# Imaging atomic-scale surface structure and superconducting gap of the misfit layered compound $(SnS)_{1.17}NbS_2$

Dongming Zhao, Shamarita Deb, Lichen Ji, Souvik Das, Paul Simon, Raúl Cardoso-Gil, Steffen Wirth, Walter Schnelle, Sushmita Chandra,* Jian-Feng Ge,* Claudia Felser

*Max Planck Institute for Chemical Physics of Solids, 01187 Dresden, Germany*

*Correspondence: Sushmita.Chandra@cpfs.mpg.de; Jianfeng.Ge@cpfs.mpg.de

## Abstract

Superlattices of van der Waals-coupled two-dimensional materials offer a platform for engineering electronic band structure and fostering emergent quantum states. Misfit layered compounds provide naturally assembled superlattices, in which crystallographically distinct two-dimensional subsystems coexist within a single bulk crystal. Therein, the presence of both commensurate and incommensurate lattice directions introduces structural modulation, which might alter real-space electronic orders such as superconductivity. Here, we investigate the atomic-scale surface structure and local superconductivity of the misfit layered compound $(SnS)_{1.17}NbS_2$ using structural characterization, electrical transport, and low-temperature scanning tunnelling microscopy (STM). While X-ray diffraction and cross-sectional scanning transmission electron microscopy confirm the bulk misfit structure, STM resolves large SnS- and $NbS_2$-terminated terraces, where the surface atomic registry differs from the bulk misfit structure. Furthermore, tunnelling spectroscopy at 2 K reveals nearly identical superconducting gaps on both terminations, and the spatial variation of the gap shows no visible modulation that corresponds to the misfit lattice. These results establish an atomic-scale picture of the surface structure and superconducting states of this naturally assembled misfit heterostructure.

## Keywords

Misfit compound, Superconductivity, Superlattice, Scanning tunneling microscopy

## Introduction

Two-dimensional (2D) materials provide a versatile building block for constructing artificial heterostructures and superlattices with emerging physical properties absent in their individual constituents. Owing to the weak van der Waals interaction between adjacent layers, dissimilar 2D crystals can be assembled without the stringent lattice-matching constraints encountered in conventional epitaxial heterostructures [1]. In particular, stacking 2D lattices with a controlled twist angle or a small lattice mismatch can generate long-wavelength moiré superlattices, which can host a variety of correlated and topological states [2]. However, such artificial structures are commonly fabricated through exfoliation and delicate layer-by-layer transfer, which requires precise control of the relative orientation and interfacial environment [3,4]. Residual contamination, interfacial bubbles, local strain, and variations in twist angle can further compromise structural homogeneity, making scalability and reproducibility persistent challenges [3,4].

Misfit layered compounds, in contrast to artificial heterostructures, are naturally grown van der Waals heterostructures, providing a particularly interesting platform that combines crystallographically distinct two-dimensional subsystems within a single crystal. These misfit compounds are commonly composed of alternating structurally dissimilar layers, often a distorted rocksalt-type ***MX*** layer (***M*** = Sn, Pb, Sb, Bi, or rare-earth elements; ***X*** = S or Se) and a hexagonal $\boldsymbol{TX}_2$ layer (***T*** = Ti, V, Cr, Nb, or Ta). Their in-plane lattice parameters are incommensurate along at least one crystallographic direction [5,6], hence the name misfit. The resulting lattice mismatch produces intrinsic structural modulation and built-in strain, while the interaction between the constituent layers can further modify their electronic structures through charge redistribution, orbital hybridization, and interlayer bonding [7,8]. Such coupling may strongly alter the carrier density and low-energy electronic states compared to those in each isolated compound [7,8]. Consequently, misfit layered materials exhibit a broad range of collective and transport phenomena, including pronounced electronic anisotropy, density-wave instabilities, unusual magnetic and thermoelectric responses, and superconductivity in selected systems [5,6,8–13].

$(SnS)_{1.17}NbS_2$ is a misfit layered superconductor composed of alternating SnS and monolayer $NbS_2$ sublayers [14–17]. The square SnS and hexagonal $NbS_2$ sublattices share a common in-plane periodicity of approximately 0.575 nm along the ***b*** direction, while remaining incommensurate along the orthogonal ***a*** direction [14–16]. This directional lattice matching and mismatch are accompanied by substantial distortions of the constituent layers [14–16]. The $NbS_2$ subsystem adopts a monolayer-like 1H configuration embedded between SnS layers; the resulting layered structure is strongly anisotropic and lacks inversion symmetry [17,18]. Previous transport measurements on this material show superconductivity with a $T_c \approx 2.7$ K and pronounced anisotropy [18]. Together, these structural and superconducting properties make $(SnS)_{1.17}NbS_2$ a prototype misfit compound for examining how lattice misfit and interlayer coupling influence local electronic states.

In this work, we combine bulk structural and transport characterization with low-temperature scanning tunnelling microscopy and spectroscopy (STM/STS) to directly resolve the SnS- and $NbS_2$-terminated surfaces and probe their local superconducting states. We find that the SnS termination exhibits an atomic registry distinct from the established bulk SnS lattice, suggesting substantial surface relaxation. In addition, our tunnelling spectroscopic measurements reveal identical superconducting gaps on the SnS and $NbS_2$ terminations. Our results provide a microscopic view of the surface structure and superconductivity in this misfit layered compound.

## Results and Discussion

We first determine the bulk crystal structure of the misfit samples. Fig. 1(a) shows the crystal structure of $(SnS)_{1.17}NbS_2$, consisting of alternating SnS and $NbS_2$ sublayers stacked along the *c* axis. We obtained the room-temperature powder X-ray diffraction pattern [Fig. 1(b)] showing sharp (00l) reflections, which indicate high crystallinity and a preferred *c*-axis orientation. The plate-like crystals exhibit shiny surfaces, with typical dimensions of ~5 mm × 7 mm × 1 mm [inset of Fig. 1(b)]. Cross-sectional scanning transmission electron microscopy (STEM) directly resolves the periodic alternation of the SnS and $NbS_2$ layers without observable stacking faults [Fig. 1(c)]. The observed stacking sequence and layer morphology are consistent with previous structural studies [14–16] of bulk $(SnS)_{1.17}NbS_2$, confirming the misfit structure of our bulk crystals.

Next, we investigate the structure of the cleaved surface of these samples. Our large-scale STM image [Fig. 2(a)] reveals two regions with distinct heights and different surface morphologies. The upper (brighter) regions on the left and right are generally flat and relatively cleaner, with the dominant defects appearing as vacancy-like features accompanied by only a small number of adsorbate-like features. In contrast, the lower (darker) region in the center appears to be rougher, with a much higher density of adsorbate-like features distributed across the surface. As we cleave the crystals at cryogenic vacuum, we speculate that these adsorbate-like features are mobile sulfur atoms due to their chemical activeness.

The pronounced difference in surface morphology serves as a preliminary indicator for distinguishing the two terminations. They can be further determined by their characteristic lattice symmetries, as illustrated in Fig. 2(b). The SnS layer exhibits a square-like lattice, while the $NbS_2$ layer has a triangular lattice with sixfold symmetry. We acquired atomic-resolution STM images on both layers within the regions marked by squares in Fig. 2(a). On the lower (darker) termination, the underlying atomic corrugation is partially obscured by the high density of adsorbate-like features, but after filtering them out, a triangular lattice becomes clearly resolved [Fig. 2(d)]. Furthermore, the Fourier transform of the unfiltered image [Fig. 2(f)] exhibits a well-defined sixfold-symmetric set of Bragg peaks, with a lattice parameter of approximately 0.33 nm. The periodicity and the sixfold symmetry are consistent with the lattice constant of the $NbS_2$ sublayer. We therefore assign the lower termination to $NbS_2$. In contrast, our atomic-resolution STM image [Fig. 2(c)] on the upper (brighter) termination exhibits a square pattern, consistent with the SnS surface lattice. We therefore assign the upper termination to SnS.

Interestingly, the Fourier transform of the SnS termination [Fig. 2(e)] shows a dominant square set of Bragg peaks corresponding to a periodicity of 0.33 nm, together with a weaker set of sixfold Bragg peaks. As the latter appears identical to the Bragg peaks in Fig. 2(f), we thus associate them with the underlying $NbS_2$ lattice. By analyzing the peaks in reciprocal space in Fig. 2(e), we reconstruct the surface atomic registry of the SnS and $NbS_2$ layers in the top-view structural model [Fig. 2(b)]. We find that the smallest structural unit, marked by the blue square in Fig. 2(b) with a periodicity of 0.33 nm, is commensurate with the lattice constant of the underlying $NbS_2$ layer. One pair of these $NbS_2$-related Bragg peaks is aligned with the 0.33 nm periodicity of the SnS surface lattice, indicating that the two sublattices match along ***a*** direction. This is in stark contrast with the bulk misfit structure in Fig. 1, which shows a commensurate direction along ***b*** direction, and with a different matching periodicity of 0.57 nm [14–16]. We attribute the difference in the surface structure in the SnS termination to surface relaxation, because the SnS sublayer in the misfit compound adopts a distorted rocksalt-type configuration, which is energetically unfavorable and dynamically unstable in the absence of stabilizing constraints [19,20].

We then proceed to examine the superconducting properties of the $(SnS)_{1.17}NbS_2$ crystals. We first measure the temperature dependence of the in-plane electrical resistivity. As shown in Fig. 3(a), the bulk resistivity exhibits metallic behavior in the normal state, followed by a superconducting transition starting at approximately 2.7 K and reaching zero resistance near 2.25 K, in agreement with previous reports [17,18]. We then performed STS measurements at 2.0 K on both the SnS and $NbS_2$ terminations. As shown in Figs. 3(b) and 3(c), the spectra acquired on the two surfaces reveal very similar, well-defined superconducting gaps. We fit the spectra using the Dynes formula [21], which yields $\Delta = 0.334 \pm 0.012$ meV on the SnS termination and $\Delta = 0.331 \pm 0.014$ meV on the $NbS_2$ termination. The fit results indicate no resolvable difference in the gap size between the two terminations. Although measuring temperature dependence of the gap is beyond the current capability of our STM, a simple weak-coupling estimate based on the Bardeen–Cooper–Schrieffer (BCS) theory is consistent with the observed gap magnitude: for $T_c = 2.7$ K, the zero-temperature gap $\Delta_0 = 1.76k_BT_c = 0.41$ meV ($k_B$ being the Boltzmann constant), and the BCS temperature dependence yields $\Delta(T = 0.75T_c) = 0.32$ meV at 2.0 K.

To examine whether the lattice misfit influences superconductivity locally at the atomic scale, we performed high-resolution spectroscopy in a $4 \times 4$ nm$^2$ area [Fig. 4(a)], and fit each spectrum using the Dynes formula. Fig. 4(b) displays the spatial distribution of the gap size extracted from the fit results, which reveals moderate nanoscale variations of the superconducting gap around 0.33 meV. Although the fitted gap varies across the field of view, its spatial distribution does not show any pattern following the atomic corrugation or defects present in the topographic image Fig. 4(a). The corresponding histogram [Fig. 4(c)] is centered near 0.33 meV, consistent with the representative spectra in Fig. 3. Within the spatial range and experimental resolution of the present measurements, we therefore find no clear evidence for a lattice-modulated superconducting gap on the SnS termination.

The almost identical superconducting gaps observed on the SnS and $NbS_2$ terminations suggest that superconducting coherence extends across the constituent layers of the misfit structure. Noting that bulk 2H-$NbS_2$ is a superconductor [22] and bulk SnS is a semiconductor [23,24], one possible scenario is that superconductivity originates from the $NbS_2$ sublayers and extends into SnS sublayers through the superconducting proximity effect. As the *c*-axis coherence length of $NbS_2$ is ~ 2 nm [25] and far exceeds the interlayer distance (~0.6 nm) in $(SnS)_{1.17}NbS_2$, Cooper pairs from $NbS_2$ layers diffuse into the neighboring SnS layers and retain high phase coherence, so that a similar superconducting gap can be measured on both layers.

One particular question in misfit superconductors is whether the intrinsic lattice mismatch couples to the local superconducting order parameter through structural modulation and spatially varying interlayer registry. This possibility is particularly interesting for $(SnS)_{1.17}NbS_2$, because bulk 2H-$NbS_2$, unlike other transition-metal dichalcogenides like $NbSe_2$ or $TiSe_2$, does not exhibit a static long-range charge-density-wave (CDW) order [22,26,27]. Consequently, no complication of intertwined long-range order due to CDW [28] would exist in the misfit compound $(SnS)_{1.17}NbS_2$; any existing spatial modulation of the superconducting order parameter may stem from the intrinsic structural mismatch. However, the present gap map [Fig. 4(b)] reveals no clear modulation of the superconducting gap associated with the misfit structure at our measurement temperature. Future Josephson STM measurements at mK temperatures [29], which probe the Cooper pair density more directly than single-particle spectra [30], may provide a more sensitive test of whether the structural misfit couples to the superconducting order parameter.

## Conclusion

In summary, we investigated the surface structure and local superconducting properties of the misfit layered compound $(SnS)_{1.17}NbS_2$ by combining bulk structural characterization with low-temperature STM and STS. X-ray diffraction and cross-sectional STEM confirm the bulk misfit structure, while STM resolves distinct SnS- and $NbS_2$-terminated surfaces with markedly different morphologies and atomic lattices. On the SnS termination, the observed 0.33 nm square periodicity indicates a different misfit structure from the bulk counterpart. On the surface, the reciprocal-space registry with the underlying $NbS_2$ lattice is consistent with a substantial relaxation of the exposed SnS surface. Tunnelling spectroscopy at 2 K reveals very similar superconducting gaps of 0.33 meV on both terminations, while spatially resolved spectra on the SnS surface show slight gap variations without modulation corresponding to the misfit structure. These results establish an atomic-scale picture of the surface structure and local superconducting states of $(SnS)_{1.17}NbS_2$ and provide a basis for future studies of how interlayer coupling and structural incommensurability influence superconductivity in misfit layered compounds.

## Methods

### Crystal synthesis

High-quality single crystals of $(SnS)_{1.17}NbS_2$ were grown by chemical vapor transport. Elemental tin (Sn), niobium (Nb), and sulfur (S) powders (Alfa Aesar, 99.99%) were first reacted at 850 °C to obtain a polycrystalline precursor of $SnNbS_3$. The resulting precursor was then sealed in a quartz ampoule with $SnCl_2$ as the transport agent and subjected to a temperature gradient of 840–770 °C for 14 days. Single crystals were formed at the lower-temperature end of the ampoule.

### Crystal characterization

The room-temperature XRD pattern was recorded by mounting the powder sample on a zero-background sample holder and using a Stoe Stadi-MP diffractometer in Bragg-Brentano geometry. The diffractometer is equipped with a DECTRIS MYTHEN2 1K silicon strip detector and operated with Cu-K$\alpha_1$ radiation [$\lambda$ = 1.54056 Å, curved germanium (111) Johann-type Monochromator]. Cross-sectional lamellae of a $(SnS)_{1.17}NbS_2$ single crystal were prepared using the standard focused ion beam (FIB) lift-out procedure on a Thermo Fisher Scientific Helios 5 CX FIB. Scanning transmission electron microscopy (STEM) images were collected in high-angle annular dark-field (HAADF) configuration on a double aberration-corrected JEOL ARM300F operating at 300 kV with a 30 mrad convergence angle.

### Transport measurement

Electrical resistivity measurement was conducted using a Quantum Design Physical Property Measurement System (PPMS) in a conventional four-probe configuration. Electrical contacts were made using Pt wires and silver paste. The samples were oriented with the *c*-axis parallel to the applied magnetic field, and an excitation current of 0.5 mA was used for the measurement.

**STM measurement**

We performed STM measurements using a commercial low-temperature STM (USM1500, Unisoku Co., Ltd.) at 2.0 K. The sample was cleaved at 77 K in the ultra-high vacuum with a base pressure of 1 × $10^{-10}$ Torr. The sample was immediately transferred to STM. We used mechanically sharpened Pt/Ir tips, whose condition was verified on the Au (111) surface. STM images were acquired in constant current mode, while differential conductance spectra were obtained using a standard lock-in technique with a modulation frequency of 983 Hz.

## Availability of data and material

The datasets generated and analyzed during the current study are openly available in the Edmond research data repository at https://doi.org/10.17617/3.IMYAUM.

## Competing interests

The authors declare no competing interests.

## Funding

This work was supported by a collaboration between The Kavli Foundation, Klaus Tschira Stiftung, and Kevin Wells, as part of the SuperC collaboration. S.C. and C. F. acknowledge financial support by the Deutsche Forschungsgemeinschaft (DFG) through QUAST (project ID FOR 5249).

## Authors' contributions

J.-F.G. and S.C. conceived the experiment. S.C. conducted the material synthesis. S.C. and P.S. performed the STEM measurement. S.C. and R.C.-G. performed the XRD measurement. D.Z. and S.Deb performed the STM measurements with the help of L. J., S.Das, and S.W.. D.Z. and S.Deb analyzed the STM data. J.F.G supervised the project with support from W.S. and C.F. D.Z. wrote the manuscript with input from all authors.

## Acknowledgements

We thank A. Kreisel for fruitful discussions.

## Figures

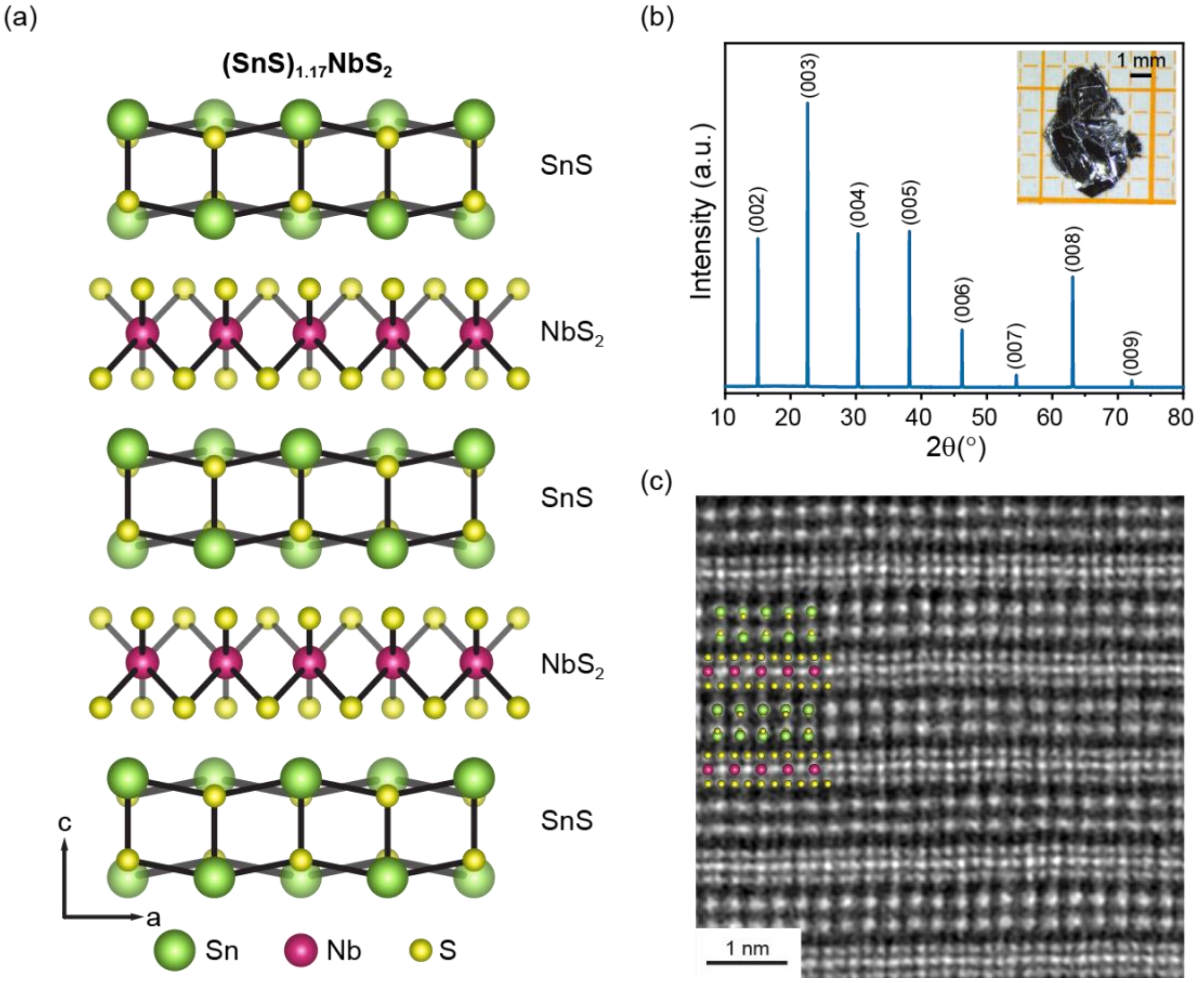

**Fig. 1 Structural characterization of $(SnS)_{1.17}NbS_2$.** (a) Crystal structure of $(SnS)_{1.17}NbS_2$ along the ***b*** axis. Sn, Nb, and S atoms are represented in green, magenta, and yellow, respectively. (b) X-ray diffraction pattern of the $(SnS)_{1.17}NbS_2$ single crystal. Inset: a photograph of the as-grown single crystal. (c) Cross-sectional STEM image along the [010] axis showing the alternating stacking of SnS and $NbS_2$ layers. The structural model is overlaid for comparison.

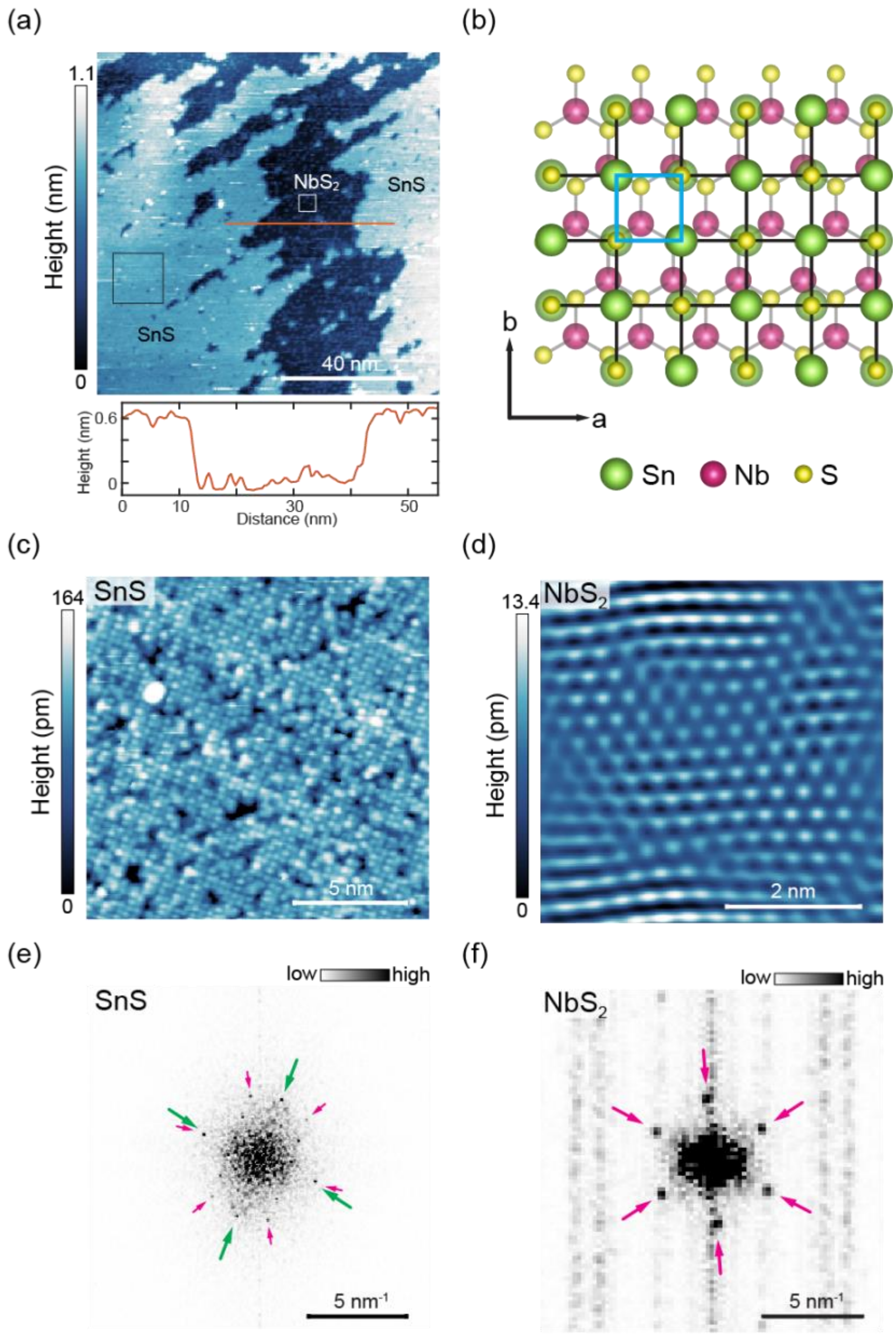

**Fig. 2 Surface structure of $(SnS)_{1.17}NbS_2$.** (a) Large-area STM topography showing adjacent SnS- and $NbS_2$-terminated regions ($V_{bias}$ = 100 mV, $I_{set}$ = 30 pA). The boxes indicate the regions enlarged in (c) and (d). The lower panel shows the height profile extracted along the orange line in the topographic image. (b) Top-view structural model illustrating the relative registry of the SnS and $NbS_2$ layers at the surface. The blue square marks the smallest square structural unit resolved experimentally on the SnS termination, corresponding to a characteristic Sn–S spacing of approximately 0.33 nm. (c) Atomic-resolution STM topography of the SnS termination ($V_{bias}$ = −100 mV, $I_{set}$ = 50 pA). (d) Fourier-filtered STM topography of the $NbS_2$ termination ($V_{bias}$ = 10 mV, $I_{set}$ = 100 pA). (e) Fast Fourier transform (FFT) of (c), showing the square Bragg pattern of the SnS termination (green arrows) and additional hexagonal Bragg peaks associated with the $NbS_2$ lattice (magenta arrows). (f) FFT of the unfiltered image of (d), with the $NbS_2$ Bragg peaks indicated by magenta arrows.

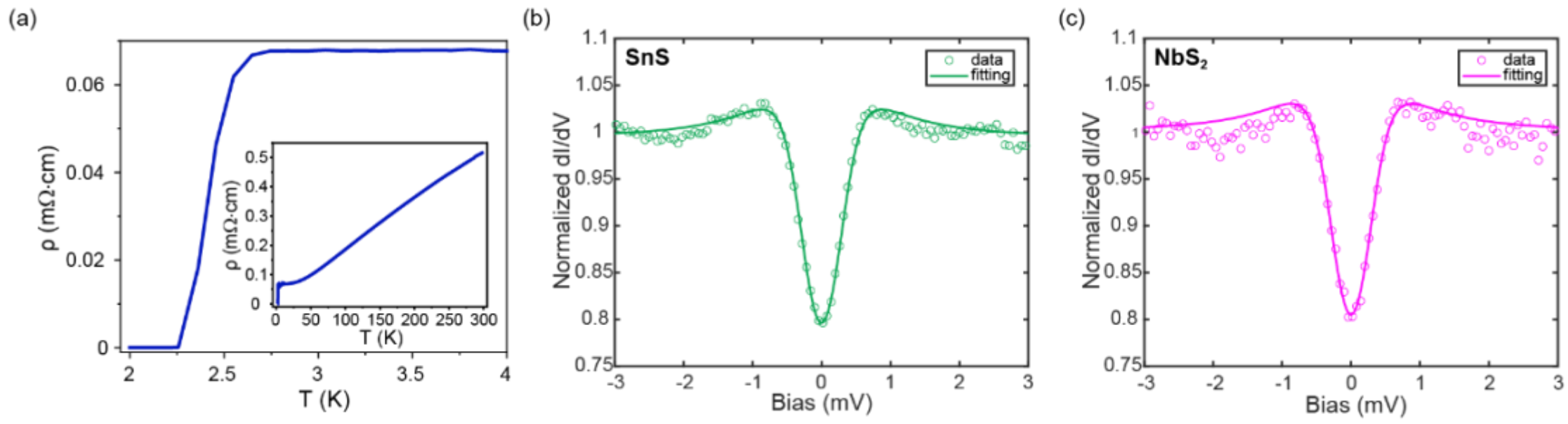

**Fig. 3 Superconductivity of $(SnS)_{1.17}NbS_2$.** (a) Temperature dependence of the resistivity near the superconducting transition with an onset of 2.7 K. Inset: resistivity over the full measured temperature range. Tunnelling spectra measured at 2.0 K (~ 0.75 $T_c$) on (b) the SnS and (c) the $NbS_2$ terminations, respectively. Solid curves show fitting by the Dynes formula. The fitted superconducting gap is $\Delta$ = 0.33 meV for both terminations. Set point: $V_{bias}$ = -3 mV, $I_{set}$ =100 pA, $V_{mod}$ = 0.06 mV.

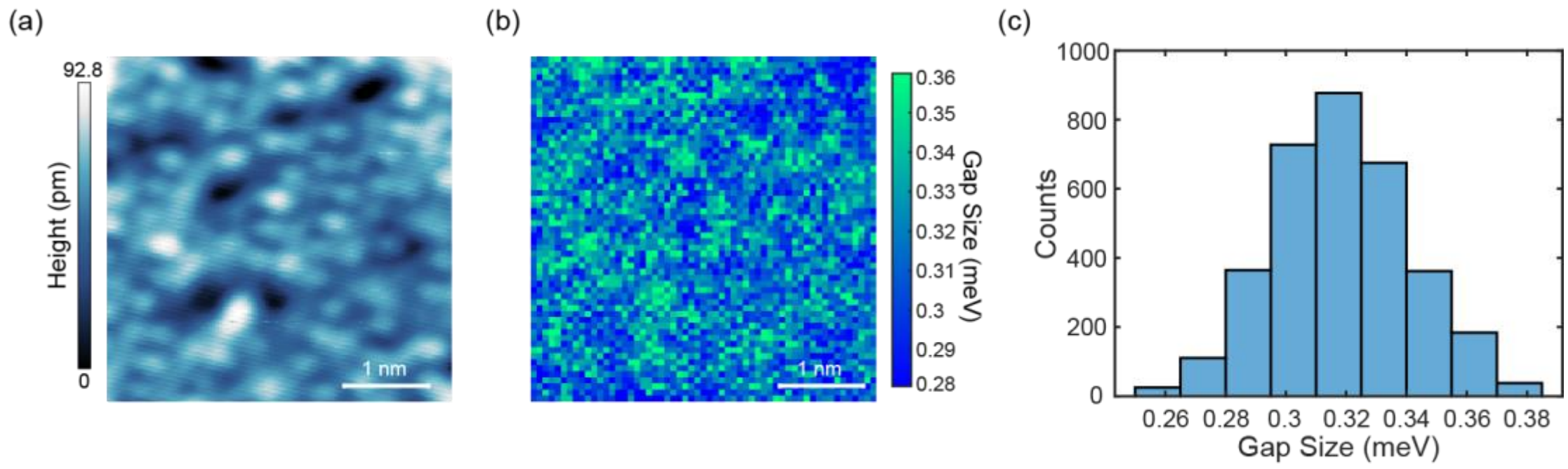

**Fig. 4 Spatial variation of the superconducting gap on the SnS termination.** (a) STM topography of the 4 × 4 $nm^2$ spectroscopic mapping area. (b) Spatial distribution of the superconducting gap Δ obtained from Dynes fits to the local tunnelling spectra. (c) Histogram of the fitted gap values in (b). Set point: $V_{bias}$ = −1 mV, $I_{set}$ = 100 pA, $V_{mod}$ = 0.03 mV.